\documentclass[prd, aps, showpacs]{revtex4}
\usepackage{latexsym}

\begin{document}
\draft

\title{Quantum Corrections to Synchrotron Radiation from Wave-Packet}
\author{Shih-Yuin Lin}
\email{sylin@phys.sinica.edu.tw}
\affiliation{Institute of Physics,
Academia Sinica, Nankang, Taipei 11529, TAIWAN }
\date{June 2003}

\begin{abstract}
We calculate the radiated energy to $O(\hbar)$ from a charged wave-packet
in the uniform magnetic field. In the high-speed and weak-field limit,
while the non-commutativity of the system reduces the classical radiation,
the additional corrections originated from the velocity uncertainty of the
wave-packet leads to an enhancement of the radiation.
\end{abstract}

\pacs{04.62.+v, 04.70.Dy}

\maketitle

\section{Introduction}
Radiation spectra such as the synchrotron radiation by accelerated
electrons had been derived for many years by applying the semi-classical
formalism\cite{qed,schw,tsai,sour,latal}. In the conventional treatments
using the relativistic quantum mechanics, while the electromagnetic field
is quantized as a field, the electrons or point-charges are quantized as
particles. Such a formalism is particularly suitable for studying the
semi-classical theories in accelerated frames\cite{unruh,dewitt,lin03b},
because the classical concepts such as the trajectory can emerge in the
form of the expectation values and help to define an accelerated frame
clearly.

To compare directly with those discussions in accelerated frame, the
expectation value of the position of the electron has to evolve. This
means that we have to choose a four-dimensional wave-packet as the quantum
state of a moving electron, though the quantum states usually used in
calculating the synchrotron radiation are ``energy" eigenstates in the
proper frame\cite{tsai, latal}. Since a single electron in the accelerator
behaves like a moving particle rather than an eigenstate, it is
interesting to check the discrepancy between our choice of quantum states
and those in conventional calculations.

In the present paper, we will work out the $O(\hbar)$ corrections to the
radiated energies emitted from a wave-packet moving in a uniform magnetic
field. The paper is organized as follows. The semi-classical formalism for
calculating the radiated energy of charges will be given in Section II.
Following this formalism, the synchrotron radiation from charges in a
uniform magnetic field will be calculated in Section III. Then we will
discuss the cases in the static limit and the high-speed, weak-field
limit. Finally, our conclusion will be given in Section IV.

\section{Radiated Energy from Accelerated Wave-Packets}

The relativistic Lorentz electron with the charge $e$ and the mass $m$ is
described by the action\cite{rohr},
\begin{equation}
  S = \int d\tau {m\over 2}v_\mu v^\mu
  +\int d^4 x \sqrt{-g}\left[-{1\over 4}F_{\mu\nu}F^{\mu\nu}
  + j_\mu (x) A^\mu(x)\right] ,\label{actEM}
\end{equation}
where $v^\mu \equiv dy^\mu(\tau)/d\tau$ is the proper velocity,
$F_{\mu\nu}\equiv D_\mu A_\nu -D_\nu A_\mu$ is the electromagnetic field
tensor, and the current of the point-like electron is defined by
\begin{equation}
  j_\mu(x)\equiv e\int d\tau v_\mu(\tau)
    \delta^4 (x-y(\tau))[-g]^{-1/2},\label{defJ}
\end{equation}
with the conservation law $j^\mu{}_{;\mu} =0$. Below we use the Cartesian
coordinate $d\tau^2 = -dt^2+dx^2+dy^2+dz^2$ and natural units $c=\hbar=1$
for convenience but keep $\hbar$ in our expressions.

In quantum electrodynamics, the energy radiated from a charged particle
via photons is the photon energy $\hbar\omega$ multiplied by the
transition probability ${\cal I}$ as\cite{qed, tsai}
\begin{equation}
  {\cal E}
  =\int{d^3 k\over (2\pi)^3}\hbar\omega{d^2{\cal I}\over d\omega d\Omega}.
\end{equation}
From the interacting action $S_{\rm int}\sim \int j_\mu A^\mu$ in
$(\ref{actEM})$, the transition probability ${\cal I}$ up to the first
order of perturbation is given by
\begin{equation}
{\cal I}= {1\over\hbar^2}{\rm Re}\left<\left|\int d^4x\sqrt{-g}
    j_\mu (x)A^\mu(x)\right|^2\right> .
\end{equation}
The quantum states for $j_\mu$ could be represented by Klein-Gordon wave
functions $\phi$, or some linear combinations of them, in the
semi-classical treatment. The trajectory of the charge in our
semi-classical theory is thus related to the expectation values of the
operators for the motion of the source.

To the first order approximation, since $A_\mu$ and $j_\mu$ are independent
dynamical variables in the free field theory, we choose the quantum state
of the system to be the direct product of the vacuum state for $A_\mu$
and the quantum states for $j_\mu =e v_\mu$. Therefore,
\begin{eqnarray}
{\cal I}&=&{e^2\over \hbar^2}{\rm Re}\int d\tau d\tau'\left< y|v_\mu(\tau)
    \left< 0|A^\mu(y(\tau))A^\nu(y(\tau'))|0\right>v_\nu(\tau')|y\right>.
\end{eqnarray}
Suppose the laboratory observer is in Minkowski frame. What the laboratory
observer can record are thus Minkowski photons. Let us take the initial
state for photons being the Minkowski vacuum $|\left.0_M \right>$, which
is defined by the creation and annihilation operators in Minkowski space.
Accordingly, the vector field could be decomposed into
\begin{equation}
  A^\mu (x) = \sum_{\lambda =1,2}\int {d^3 k\over (2\pi )^3}
  \left[ a_\lambda({\bf k})\epsilon_\lambda^\mu({\bf k})\chi(\omega)
  e^{i{\bf k}\cdot{\bf x}}+ a^\dagger_\lambda({\bf k})
    \epsilon_\lambda^{*\mu}({\bf k})\chi^*(\omega)
  e^{-i{\bf k}\cdot{\bf x}}\right],
\end{equation}
where $\chi(\omega)=\sqrt{\hbar/2\omega}e^{-i\omega t}$ with $\omega =
|{\bf k}|$, such that
\begin{equation}
  {\cal I}={e^2\over 16\pi^3\hbar}{\rm Re}\int d\omega d\Omega\omega
    \int d\tau d\tau' \left<y|\right.v_\mu(\tau) e^{i k_\alpha
    y^\alpha(\tau)}e^{-ik_\alpha y^\alpha(\tau')}
    v^\mu(\tau')\left.|y\right> , \label{probabE}
\end{equation}
hence the spectral and angular distribution of the radiated energy reads
\begin{equation}
  {d^2{\cal E}\over d\omega d\Omega} = {e^2 \omega^2\over 16\pi^3}
   {\rm Re}\int d\tau d\tau' \left<y|\right.v_\mu(\tau)
   e^{i k_\alpha y^\alpha(\tau)}e^{-ik_\alpha y^\alpha(\tau')}
   v^\mu(\tau')\left.|y\right>.\label{radE}
\end{equation}
Note that the above $\left< A^\mu A^\nu\right>$ are functions of $y_\mu
(\tau)$, which do not commute with $v_\mu = (-i\hbar\partial_\mu -eA_\mu)
/m$. This non-commutativity is understood as an origin of the quantum
correction\cite{qed}. If we ignore the non-commutativity of the operators,
the classical formula will be explicitly obtained by substituting classical
current $v^\mu$ into Eq.$(\ref{radE})$\cite{tsai}\cite{jack}.

Let us concentrate on the expectation value in the differential radiated
energy $(\ref{radE})$, namely,
\begin{equation}
 {\cal A} \equiv \left<v_\mu(\tau)e^{i k_\alpha y^\alpha(\tau)}
   e^{-ik_\alpha y^\alpha(\tau')}v^\mu(\tau')\right> .
\end{equation}
Denoting the fluctuations $\delta y^\mu(\tau)=y^\mu(\tau)- \left<y^\mu
(\tau) \right>$ and $\delta v^\mu(\tau)=v^\mu (\tau)-\left< v^\mu(\tau)
\right>$ and expanding ${\cal A}$ in terms of $\delta y^\mu$ and $\delta
v^\mu$, one has
\begin{equation}
{\cal A}=\left[\left< v_\mu (\tau)\right>\left< v^\mu (\tau')\right>+
  \sum_{i=1}^3 f_i\right] e^{ik_\alpha\left[\left<y^\alpha(\tau)\right>-
  \left<y^\alpha(\tau')\right>\right]}+O(\hbar^2),\label{calA}
\end{equation}
in which the $O(\hbar)$ corrections $f_i$ are
\begin{eqnarray}
  f_1 &\equiv&\left<\delta v_\mu (\tau)\delta v^\mu (\tau')\right>,
    \label{f1}\\
  f_2 &\equiv& -{k_\mu k_\nu\over 2} \left[\left<\delta y^\mu(\tau)
    \delta y^\nu(\tau)\right>+ \left<\delta y^\mu(\tau')\delta y^\nu(\tau')
    \right>-2\left<\delta y^\mu(\tau)\delta y^\nu(\tau')\right>\right]
    \left< v_\rho (\tau)\right>\left< v^\rho(\tau')\right> ,\label{f2}\\
  f_3 &\equiv& ik^\nu\left\{\left<v^\mu(\tau')\right>\left<\delta v_\mu
    (\tau )\left[ \delta y_\nu(\tau)-\delta y_\nu(\tau')\right]\right> +
    \left<v_\mu(\tau)\right>\left<\left[\delta y_\nu(\tau)-
    \delta y_\nu(\tau')\right]\delta v^\mu(\tau')\right>\right\}.\label{f3}
\end{eqnarray}
Because of the dependence on $k^\mu$, one expects that $f_2$ and $f_3$
will become more important than $f_1$ in the short wave-length regime.

If we choose the quantum state $\left.|y\right>$ to be a four-dimensional
wave-packet centered at some trajectory $\left<y_\mu(\tau)\right>$ with
the four-velocity $\left<v_\mu (\tau)\right>=d\left<y_\mu(\tau)\right>/
d\tau$, then the first term in the bracket of Eq.$(\ref{calA})$ gives the
classical radiated energy by the charge moving in the trajectory
$\left<y_\mu(\tau) \right>$, while the $f_1$ term might be interpreted as
the Unruh effect on the charged current\cite{dewitt}.

To calculate the $f_i$'s, one needs the Hamiltonian for the charge motion.
From the sector of the charge motion in the action $(\ref{actEM})$, the
conjugate momentum for $y^\mu$ is $p^\mu = mv^\mu + eA_{\rm in}^\mu$.
Hence the ``Hamiltonian" with respect to the proper time $\tau$ reads
\begin{equation}
  {\cal H} = {1\over 2m}(p_\mu -eA^{\rm in}_\mu)(p^\mu -eA_{\rm in}^\mu),
  \label{hami}
\end{equation}
after a Legendre transformation. The quantum mechanics for the Lorentz
electron is given by the equal-proper-time commutation relation
\begin{equation}
  \left[ y_\mu(\tau), p_\nu(\tau) \right] = i\hbar g_{\mu\nu}.
\end{equation}
It follows that
\begin{equation}
  \left[v^\mu(\tau),v^\nu (\tau)\right] =i\hbar {e\over m^2}F_{\rm in}^{\mu\nu},
\label{noncomVV}
\end{equation}
so $v_\mu(\tau)$ and $v^\nu(\tau')$ do not commute in the presence of the
background electromagnetic field. This implies the uncertainty relation,
\begin{equation}
  \sqrt{\left<\delta v_\mu^2(\tau)\right>
  \left<\delta v_\nu^2(\tau)\right> } \ge
  {\hbar\over 2m^2}\left|eF^{\rm in}_{\mu\nu}\right|. \label{vuncert}
\end{equation}

\section{Synchrotron Radiation in Uniform Magnetic Field}

For a charge moving in a uniform magnetic field $B^3=F_{\rm in}^{12}=H$,
the Heisenberg equation of motion gives the evolution of the operators as
follows,
\begin{eqnarray}
  v_1(\tau) &=& \hat{v}_1\cos\omega_0\tau +\hat{v}_2\sin\omega_0\tau,\\
  v_2(\tau) &=& \hat{v}_2\cos\omega_0\tau -\hat{v}_1\sin\omega_0\tau,\\
  v_0(\tau) &=& \hat{v}_0,\;\; v_3(\tau) = \hat{v}_3,\\
  y_1(\tau) &=& \hat{y}_1+{\hat{v}_1\over\omega_0}\sin\omega_0\tau -
     {\hat{v}_2\over\omega_0}(\cos\omega_0\tau -1),\\
  y_2(\tau) &=& \hat{y}_2+{\hat{v}_2\over\omega_0}\sin\omega_0\tau +
     {\hat{v}_1\over\omega_0}(\cos\omega_0\tau -1),\\
  y_0(\tau) &=& \hat{y}_0+\hat{v}_0\tau, \;\;
  y_3(\tau) = \hat{y}_3+\hat{v}_3\tau,
\end{eqnarray}
where $\omega_0=eH/m$, $\hat{v}_\mu\equiv v_\mu(0)$ and $\hat{y}_\mu\equiv
y_\mu(0)$. Substituting above operators into $(\ref{f1})$-$(\ref{f3})$,
the $f_i$'s can be worked out straightforwardly, for instance,
\begin{eqnarray}
  f_1 &=& \left<v_\mu(\tau)v^\mu(\tau')\right>-
  \left<v_\mu(\tau)\right>\left<v^\mu(\tau')\right> \nonumber\\
  &=& -\left<\delta\hat{v}_0{}^2\right>+\left<\delta\hat{v}_3{}^2\right>+
  \left(\left<\delta\hat{v}_1{}^2\right>+\left<\delta\hat{v}_2{}^2\right>
  \right)\cos\omega_0(\tau-\tau')-  \nonumber\\ & &
  \left[\hat{v}_1,\hat{v}_2\right]\sin\omega_0(\tau-\tau').
\end{eqnarray}
Note that the last term in the above equation is an odd function of
$(\tau-\tau')$, and the techniques of integrations would be different from
those for even functions of $(\tau-\tau')$. Below we call the radiated
energies contributed by the odd and even functions of $(\tau-\tau')$ in
$f_i$'s as the ``odd" and ``even" part of the radiated energy
respectively.

Let us consider the initial condition $\left<\hat{v}^\mu\right>= (\gamma,
\gamma v,0,0)$ with $\gamma^{-1}\equiv\sqrt{1-v^2}$ while
$\delta\hat{v}_3$ is zero. Then the classical radiated energy, contributed
by $\left<v_\mu(\tau)\right>\left<v^\mu(\tau ')\right>$ in $(\ref{calA})$,
follows immediately as
\begin{equation}
{\cal P}_0={e^2\omega_0^2\over 4\pi}{2\over 3}\gamma^2 v^2 . \label{P0cl}
\end{equation}
Employing the same techniques given in Ref.\cite{tsai}, one obtains the
radiated power per unit coordinate time up to $O(\hbar)$, ${\cal P}\equiv
{\cal E}/\int \gamma d[(\tau+\tau')/2] $, coming from the ``even" part of
$f_i$'s:
\begin{eqnarray}
  {\cal P}^{\rm even}_{f_1} &=&{e^2\omega_0^2\over 4\pi}\left\{
   {1-\gamma^2\over 3}\left<\delta \hat{v}_0^2\right>+
   {2+\gamma^2\over 3}\left(\left<\delta \hat{v}_1^2\right>+
   \left<\delta \hat{v}_2^2\right>\right)\right\},\label{Peven1}\\
  {\cal P}^{\rm even}_{f_2} &=&{e^2\omega_0^2\over 4\pi} \left\{
   \left(3-7\gamma^2+4\gamma^4\right)\left<\delta\hat{v}_0^2\right>
   +\left({20\over 3}-11\gamma^2+4\gamma^4\right)
   \left<\delta\hat{v}_1^2\right>
   +\left(-{4\over 3}+\gamma^2\right)\left<\delta\hat{v}_2^2\right>
   \right\},\label{Peven2}\\
  {\cal P}^{\rm even}_{f_3} &=&{e^2\omega_0^2\over 4\pi}\left\{
    \left({-2+14\gamma^2\over 3}-4\gamma^4\right)
    \left<\delta\hat{v}_0^2\right>
    +\left({-16+4\gamma^2\over 3}+4\gamma^4\right)
    \left<\delta \hat{v}_1^2\right>\right\},\label{Peven3}
\end{eqnarray}
all of which are proportional to the velocity uncertainty of the
wave-packet at $\tau=0$. The odd part of the radiated energies are
\begin{eqnarray}
  {\cal E}^{\rm odd}_{f_i} = {e^2\over 16\pi^3}&{\rm Re}&\int_0^{2\pi}
  d\varphi\int_0^\pi d\theta \sin\theta \int_0^\infty d\omega\omega^2
    \int d\tau d\tau' f_i^{\rm odd}e^{-i\omega\gamma(\tau-\tau')}
    \times\nonumber\\ &&\exp i{\omega\over\omega_0}\gamma v\sin\theta
    \left[\cos\varphi\left(\sin\omega_0\tau-\sin\omega_0\tau'\right)
    \sin\varphi\left(\cos\omega_0\tau-\cos\omega_0\tau'\right)\right] ,
\label{Eoddf}
\end{eqnarray}
where
\begin{eqnarray}
  f^{\rm odd}_1 &=& -i\hbar {\omega_0\over m}\sin\omega_0(\tau-\tau'),\\
  f^{\rm odd}_2 &=& -i\hbar {\omega^2\over 2 m \omega_0}\sin^2\theta
    \left[1-v^2\cos\omega_0(\tau-\tau')\right]
    \left[\omega_0(\tau-\tau')-\sin\omega_0 (\tau-\tau')\right],\\
  f^{\rm odd}_3 &=& \hbar{\omega\over m}v\sin\theta\sin\omega_0(\tau-\tau')
    \left[\cos\varphi(\sin\omega_0\tau -\sin\omega_0\tau')
          \sin\varphi(\cos\omega_0\tau -\cos\omega_0\tau')\right] .
\end{eqnarray}
They are coming from the non-commutativity $(\ref{noncomVV})$ of the
system and independent of the choice of the quantum state. Unfortunately,
we did not succeed to evaluate ${\cal E}^{\rm odd}_{f_i}$ for general $v$.
In the following we will discuss the cases in the static limit and the
high-speed, weak-field limit.

\subsection{Static Limit}

In the static limit $v=0$, $\gamma\to 1$, the classical radiated power
${\cal P}_0$ in Eq.$(\ref{P0cl})$ vanishes. The sum of the even radiated
power $(\ref{Peven1})$-$(\ref{Peven3})$ now reads
\begin{equation}
  \lim_{v\to 0}\; {\cal P}^{\rm even}_{f_1}+ {\cal P}^{\rm even}_{f_2}+
  {\cal P}^{\rm even}_{f_3} = {e^2\omega_0^2\over 4\pi} {2\over 3}\left(
   \left<\delta\hat{v}_1^2\right> +\left<\delta\hat{v}_2^2\right>\right),
\label{Pevenv0}
\end{equation}
in which ${\cal P}^{\rm even}_{f_3}|_{v\to 0}$ is actually zero. If the
wave-packet is identical to the ``ground state" wave function, $\psi$ in
Eq.$(\ref{wavefn})$ with $v=0$, the above even part of the radiated power
with the velocity uncertainty $(\ref{minunc})$ will be exactly cancelled
by the sum of the odd radiated power,
\begin{equation}
  \lim_{v\to 0} \; {\cal P}^{\rm odd}_{f_1}+{\cal P}^{\rm odd}_{f_2}+
   {\cal P}^{\rm odd}_{f_3} = {e^2\omega_0^2\over 4\pi}
   {\hbar|\omega_0|\over m}\left(-1 +{1\over 3}+0\right) .
\label{Poddv0}
\end{equation}
Hence, as we expected, a static charge in the ground state does not
radiate any photon.

Excited eigenstates are also static. Sandwiched by some excited
eigenstate, the expectation values of the velocity $\left<
v_1(\tau)\right>$ and $\left< v_2(\tau)\right>$ will be zero, while the
velocity uncertainties $\left< \delta\hat{v}_1^2 \right>$ and $\left<
\delta\hat{v}_2^2\right>$ will be greater than $\hbar|\omega_0| /2m$ in
general. This makes the sum of Eq.$(\ref{Pevenv0})$ and $(\ref{Poddv0})$
positive, which implies that a static charge in some excited eigenstate
will radiate photon with the same power as the classical radiation by a
charge moving in the same background with a speed $\tilde{v}$ defined by
\begin{equation}
  {m\over 2(1-\tilde{v}^2)}\equiv\left<{\cal H}\right>_{n,L_z}-
  \left<{\cal H}\right>_{0,0},
\end{equation}
according to $(\ref{eigenE})$. Here $n$ is the principal quantum number
and $L_z$ is the angular momentum parallel to the uniform magnetic field.
This non-zero radiated energy actually corresponds to the spontaneous
emission of the system\cite{schiff,low}.  Such an energy dissipation of
the wave-packet continues until the system finally reaches the ground
state.

\subsection{High-Speed, Weak-Field Limit}

In the high-speed limit $v\to 1$ or $\gamma\gg 1$ with the weak background
field $|eH|/m^2\ll 1$, the radiated photons are mainly in the short
wave-length regime. 
The radiation is concentrated in a narrow cone in this limit, such that
only a small part of the trajectory is effective in producing the
radiation observed in a given direction. Explicitly, only the time
interval $(\tau-\tau')$ with $[\omega_0 (\tau-\tau')]^2 \sim \gamma^{-2} =
1-v^2 \approx 2(1-v)$ is effective\cite{tsai}. According to this
observation, one keeps the lowest order of $(\tau-\tau')$, for example,
\begin{eqnarray}
  \tau -\tau' - {2v\over\omega_0}\sin {\omega_0\over 2}(\tau -\tau')
    &\approx& (1-v)(\tau-\tau') +{\omega_0^2\over 24}(\tau-\tau')^3,\\
  \tau -\tau' + {2v\over\omega_0}\sin {\omega_0\over 2}(\tau -\tau')
    &\approx& 2(\tau-\tau') ,
\end{eqnarray}
in $f^{\rm odd}_i$'s to simplify the integration. By noting that
\begin{equation}
  \int d\omega\omega^2\int_{-\infty}^{\infty} dz z\sin\omega z =0 ,
\end{equation}
the odd part of the $O(\hbar)$ radiated power becomes
\begin{eqnarray}
  \lim_{v\to 1}{\cal P}^{\rm odd}_{f_1}+{\cal P}^{\rm odd}_{f_2}+
   {\cal P}^{\rm odd}_{f_3}  &=& {e^2\omega_0^2\over 4\pi}
   {\hbar|\omega_0|\over m}\left( 0+{5\sqrt{3}\over 24}\gamma^3
    -{5\sqrt{3}\over 2}\gamma^3 + O(\gamma)\right)\nonumber\\ &\approx&
    {e^2\omega_0^2\over 4\pi}\left( -{55\sqrt{3}\over 24}\right)
   {\hbar|\omega_0|\over m}\gamma^3 .\label{P5524}
\end{eqnarray}
This is the well-known result in the semi-classical
electrodynamics\cite{qed,schw}. The negative sign contributed by ${\cal
P}^{\rm odd}_{f_3}$ indicates that this quantum effect tends to reduce the
classical radiation. It can be interpreted as the absorption of the system
from the background magnetic field\cite{CTan}.

The even part of the $O(\hbar)$ radiated power $(\ref{Peven1})$-$(
\ref{Peven3})$ in the same limit is, however, positive:
\begin{equation}
  \lim_{v\to 1} {\cal P}^{\rm even}_{f_1}+ {\cal P}^{\rm even}_{f_2}
  +{\cal P}^{\rm even}_{f_3} = {e^2\omega_0^2\over 4\pi}
  \left(8\left<\delta\hat{v}_1^2\right>\gamma^4  + O(\gamma^2)\right) .
  \label{Pevenhi}
\end{equation}
If one further chooses the initial wave-packet as the Gaussian wave-packet
$(\ref{wavefn})$ with the velocity uncertainty $(\ref{minunc})$, the
negative result in $(\ref{P5524})$ would be overpowered by
$(\ref{Pevenhi})$ as $\gamma\gg 1$, and the radiated power from the
Gaussian wave-packet $(\ref{wavefn})$ at $\tau=0$ would be enhanced by the
quantum effect. This enhancement indicates that the energy dissipation via
radiation would cause a dispersion of the wave-packet in addition to the
decrease of the radius of its circular motion. By noting that
$\left<\delta\hat{v}_1^2\right> = m^{-2}\left<(\delta p_1- eH \delta
y_2/2)^2 \right>|_{\tau=0}$, and $y_2$-axis is parallel to the radial
direction of the circular motion at $\tau=0$, one sees that the quantum
correction $(\ref{Pevenhi})$ tends to squeeze the Gaussian wave-packet
$(\ref{wavefn})$  along the radial axis. This is due to the tidal force
induced by the different dissipation rates for different portions of the
wave-packet with different radius in the circular motion. In our system,
the particle in an outer orbit has a greater acceleration, hence a larger
radiated power, and falls instantaneously faster than the particle in an
inner orbit.

A typical synchrotron radiation X-ray source has $H \sim 1$ Tesla produced
by its bending magnets and $\gamma \sim 10^4 $ for the electrons in its
storage ring\cite{srrc}. This corresponds to a quantum correction ${\cal
P}^{\rm odd}$ in $(\ref{P5524})$ with the magnitude about $10^{-6}$ of the
classical radiation, well under the typical energy spread of the electron
beam with the order of $10^{-3}$. But ${\cal P}^{\rm even}$ in
$(\ref{Pevenhi})$ for the Gaussian wave-packet $(\ref{wavefn})$ is about
$\gamma$ times the magnitude of ${\cal P}^{\rm odd}$, thus, $10^{-2}$ of
the power of the classical synchrotron radiation. This is quite above the
background noise and should be observed in the accelerator if it exists.

Nevertheless, because of the presence of those complicated quantum
corrections in $(\ref{Peven1})$-$(\ref{Peven3})$ for every $\gamma$, the
wave-packet is distorted throughout the accelerating process. If an
electron in a uniform magnetic field is accelerated from its ground state,
the final configuration in the storage ring is not likely to be a Gaussian
wave packet similar to $(\ref{wavefn})$. Rather, a wave-packet giving the
lowest quantum corrected synchrotron radiation is expected.

Observing that, if the initial wave-packet is prepared with $\left<\delta
\hat{v}_1^2 \right>\sim\gamma^{-1}$, $\left< \delta\hat{v}_2^2 \right>
\sim \gamma$ and $\left<\delta\hat{v}_0^2\right> \lesssim \gamma^{0}$,
then the leading order of the even radiated power
\begin{equation}
  \lim_{v\to 1} {\cal P}^{\rm even}_{f_1}+
  {\cal P}^{\rm even}_{f_2}+{\cal P}^{\rm even}_{f_3} \approx
  {e^2\omega_0^2\over 4\pi}\left[ 8\gamma^4\left<\delta\hat{v}_1^2\right>+
  {4\over 3}\gamma^2\left<\delta \hat{v}_2^2 \right> \right]
\label{minPeven1}
\end{equation}
is proportional to $\gamma^3$, just the same order of magnitude as the odd
part of the radiated power $(\ref{P5524})$. Further, from the uncertainty
relation $(\ref{vuncert})$, one has
\begin{eqnarray}
  8\gamma^4\left<\delta\hat{v}_1^2\right>+{4\over 3}\gamma^2
  \left<\delta \hat{v}_2^2 \right> \ge  2\sqrt{{32\over 3}\gamma^6
  \left<\delta \hat{v}_1^2 \right>\left<\delta\hat{v}_2^2 \right>}
  \ge {\hbar|\omega_0|\over m}{4\sqrt{6}\over 3}\gamma^3 ,
\label{minPeven2}
\end{eqnarray}
where the equality occurs when $8\gamma^4\left<\delta\hat{v}_1^2\right> =
4\gamma^2\left<\delta \hat{v}_2^2 \right>/3$ while the wave-packet has the
minimal uncertainty. Hence the most probable configuration of the
wave-packet in the accelerator is the one with $\left<\delta \hat{v}_1^2
\right> = \hbar |\omega_0|/2\sqrt{6} m \gamma$ and $\left<\delta
\hat{v}_2^2 \right> = \hbar \sqrt{6}|\omega_0| \gamma/2 m$, which is
highly squeezed in the radial direction and looks far from the wave
function $(\ref{wavefn})$, the ``ground state" wave function in the
co-moving frame.

Substituting $(\ref{minPeven2})$ into $(\ref{minPeven1})$, one finds that
the minimum value of ${\cal P}^{\rm even}$ is so close to the $|{\cal
P}^{\rm odd}|$ in $(\ref{P5524})$ that the total $O(\hbar)$ quantum
correction to the synchrotron radiation is not detectable in today's
synchrotron radiation X-ray sources. The lowest total $O(\hbar)$ quantum
correction ${\cal P}^{\rm even}+ {\cal P}^{\rm odd}$ is again negative,
with the magnitude about one fifth of the well-known result ${\cal P}^{\rm
odd}$ in $(\ref{P5524})$.

\section{Conclusion}

The $O(\hbar)$ correction to the synchrotron radiation from a Lorentz
electron moving in a uniform magnetic field has been calculated in the
static limit and the high-speed, weak-field limit. While the conventional
calculations are mainly focused on the eigenstates for the electron, we
consider the four-dimensional wave-packet state centered at the classical
trajectory of the electron. We found that the velocity uncertainty of the
wave-packet gives an additional quantum correction to the radiation.

In the static limit, the positive corrections from the velocity
uncertainty or the width of the wave-packet cancel the negative
corrections from the non-commutativity of the system, so that the radiated
energy from the ground state wave-packet vanishes. For the excited
eigenstates, the velocity uncertainty of the wave-packet induces a
positive radiated energy corresponding to the spontaneous emission.

In the high-speed and weak-field limit, we recovered the well-known
negative result as a part of our $O(\hbar)$ correction. We found that the
velocity uncertainty of the wave-packet may overpower the conventional
result and make the total $O(\hbar)$ correction positive in some cases.
This indicates that not only the circular trajectory but also the
configuration of the moving wave-packet are unstable in the background of
uniform magnetic field. The most stable configuration of the wave-packet
has the tangential velocity uncertainty $\left<\delta\hat{v}_1^2 \right>
=\hbar |\omega_0|/2\sqrt{6}m\gamma$ and the radial velocity uncertainty
$\left<\delta\hat{v}_2^2\right>=\hbar \sqrt{6}|\omega_0|\gamma /2m$, which
contributes the lowest $O(\hbar)$ quantum correction to the radiated
energy.

\begin{acknowledgments}
I wish to thank S. C. Lee, J. -P. Wang, T. F. Jiang, and K.-W. Ng for many
helpful comments. I also thank C. I Kuo for illuminated discussions.
\end{acknowledgments}

\begin{appendix}
\section{Gaussian Wave-Packet}

To recover the circular motion of a classical charge in a uniform magnetic
field, $F^{12}_{\rm in}=H$, it is convenient to choose the symmetric gauge
such that $A_\mu =(0,-H y_2 /2,H y_1/2, 0)$\cite{flugge}. Under this
choice the Hamiltonian $(\ref{hami})$ reads
\begin{equation}
  {\cal H} = {1\over 2m} p_\mu p^\mu +{eH\over 2m}
    (p_1 y_2-p_2 y_1) + {e^2H^2\over 8m}( y_1^2 + y_2^2) .
\label{Hdeltay}
\end{equation}
The eigenstates for the charge in a uniform magnetic field in
$y^3$-direction are therefore equivalent to those for a two-dimensional
simple harmonic oscillator.

The simplest four-dimensional wave-packet moving in the classical
trajectory $\left<y^\mu (\tau)\right>=y_{cl}^\mu = (\gamma\tau, (\gamma
v/\omega_0)\sin \omega_0 \tau, (\gamma v/\omega_0)\cos\omega_0\tau , 0)$
with the minimal ``energy" is the Gaussian wave-packet described by the
wave function,
\begin{equation}
  \psi(x)  = N e^{-{i\over\hbar}{\cal E}\tau}
     e^{{i\over\hbar}p_3 y^3}\sqrt{1\over\tau+i\sigma}
  \exp\left[-{im\over 2\hbar}{\delta t^2\over\tau+i\sigma}
    -{i\over \hbar}m\gamma \delta t -{|eH|\over 4\hbar}\rho^2 +
   {i m\over 2\hbar}\gamma v \rho \cos (\phi+ \omega_0\tau)\right],
\label{wavefn}
\end{equation}
where $\delta t\equiv t-\gamma\tau$, $N$ is the normalization factor,
$\sigma$ is a constant determined by initial conditions, and the local
polar coordinate $(\rho,\phi)$ is defined by
\begin{equation}
  \rho e^{i\phi} = x-{\gamma v\over\omega_0}\sin\omega_0 \tau+
    i \left(y -{\gamma v\over\omega_0}\cos\omega_0 \tau\right) .
\end{equation}
The constant ${\cal E}$ in $\psi$ is
\begin{equation}
 {\cal E}= {m\over 2}\gamma^2 + {\hbar|\omega_0|\over 2},\label{eigenE}
\end{equation}
where the first term $m\gamma^2/2$ is up to the classical level, and we
had set $p_3$ to zero at $\tau=0$. One can verify that the above wave
function $(\ref{wavefn})$ does satisfy the Schr\"odinger equation
$(i\hbar\partial_\tau - {\cal H})\psi=0$.

When $v=0$, $\psi$ is the ``ground state" of this harmonic-oscillator
system, while for $v>0$, $\psi$ is a superposition of the excited states
for the system, though $\psi$ looks the same as the ``ground state" wave
function in the co-moving frame. By noting that $\left<v^\mu\right>=
v_{cl}^\mu = (\gamma,\gamma v\cos \omega_0\tau, -\gamma v\sin\omega_0\tau
,0)$, the expectation values of $v^2_1$ and $v^2_2$ for the wave-packet
$\psi$ in $(\ref{wavefn})$ can be written as
\begin{eqnarray}
  \left<v_1^2\right>&=&\int d^4 x\left|{1\over m}\left({\hbar\over i}
    \partial_x +{eH\over 2} y\right)\psi\right|^2 =
    (v_{cl}^1)^2 +{\hbar|\omega_0|\over 2m},\label{deltav1}\\
  \left<v_2^2\right>&=&\int d^4 x\left|{1\over m}\left({\hbar\over i}
    \partial_y -{eH\over 2} x\right)\psi\right|^2 =
    (v_{cl}^2)^2 +{\hbar|\omega_0|\over 2m},\label{deltav2}
\end{eqnarray}
for all $\tau$. Hence the velocity uncertainties for the wave-packet at
$\tau =0$ are
\begin{equation}
 \left<\delta\hat{v}_1^2\right> =\left<\delta\hat{v}_1^2\right>
  ={\hbar|\omega_0|\over 2m}.  \label{minunc}
\end{equation}
Besides, 
if $\sigma$ is a real number, one has
\begin{equation}
   \left<\hat{v}_0^2\right> =\int d^4 x\left|{1\over m}{\hbar\over i}
    \partial_t \psi\right|^2 = \gamma^2+{\hbar\over 2m\sigma},\\
\end{equation}
such that $\left<\delta\hat{v}_0^2\right>=\hbar /2m\sigma$. The value of
$\left<\delta\hat{v}_0^2\right>$ or $\sigma$ is not important either in
the static limit or in the high-speed and weak-field limit.
\end{appendix}


\begin{thebibliography}{99}

\bibitem{qed} V. B. Berestetskii, E. M. Lifshitz and L. P. Pitaevskii,
Quantum Electrodynamics, 2nd edition, Pergamon, New York, 1982, ch X and
the reference therein.

\bibitem{schw} J. Schwinger, Proc. Nat. Acad. Sci. U.S.A. 40 (1954) 132.

\bibitem{tsai} J. Schwinger and W.-Y. Tsai, Ann. Phys. 110 (1978) 63.

\bibitem{sour} J. Schwinger, Particles, Sources and Fields, Vol 1, Addison-Wesley,
Redwood, 1970, ch 2 and 3.

\bibitem{latal} H. G. Latal and T. Erber, Ann. Phys. 108 (1977) 408.

\bibitem{unruh} W. G. Unruh, Phys. Rev. D14 (1976) 870.

\bibitem{dewitt} B. S. DeWitt, in: S. W. Hawking and W. Israel (Eds.), General
Relativity: an Einstein Centenary Survey, Cambridge Uniersity Press,
Cambridge, 1979.

\bibitem{lin03b} S.-Y. Lin, to appear in Phys. Rev. D.

\bibitem{rohr} F. Rohrlich, Classical Charged Particles, Addison-Wesley,
Redwood, 1965, ch 4, 5 and 6.


\bibitem{jack} J. D. Jackson, Classical Electrodynamics, 2nd edition, Wiley, New
York, 1975, ch 14.

\bibitem{schiff} L. I. Schiff, Quantum Mechanics, 3rd edition, McGraw-Hill,
New York, 1971, ch 11.

\bibitem{low} F. E. Low, Am. J. Phys. 29 (1961) 298.

\bibitem{CTan} C. Cohen-Tannoudji, J. Dupont-Roc and G. Grynberg,
Atom-Photon Interactions: Basic Processes and Applications, Wiley, New
York, 1992, ch IV.

\bibitem{srrc} For example, National Synchrotron Radiation Research Center
in Taiwan, http://www.nsrrc.org.tw and Advanced Photon Source at Argonne
National Laboratory, http://www.aps.anl.gov

\bibitem{flugge} S. Fl\"ugge, Practical Quantum Mechanics, Springer-Verlag,
Berlin, 1974, Vol 1, ch F.

\end{thebibliography}
\end{document}